\documentclass[aps,prb,twocolumn,showpacs,floatfix,superscriptaddress]{revtex4-2}

\usepackage[utf8]{inputenc}
\usepackage[T1]{fontenc}
\usepackage[margin=2cm]{geometry}
\usepackage{float}
\usepackage{color}
\usepackage{graphicx}
\usepackage{amsmath}
\usepackage{amssymb}
\usepackage{bbm}
\usepackage{indentfirst}
\usepackage{dcolumn}
\usepackage{soul}
\usepackage{mathtools}

\usepackage{subfigure}
\usepackage{multirow}
\usepackage{setspace}

\begin{document}

\title{Spin reorientation transition in an ultrathin Fe film on W(110) induced by Dzyaloshinsky-Moriya interactions}

\author{Balázs Nagyfalusi}
\email{nagyfalusi@phy.bme.hu}
\affiliation{Department of Theoretical Physics, Budapest University of Technology and Economics, Budafoki \'{u}t 8, H-1111 Budapest, Hungary}
\author{L\'{a}szl\'{o} Udvardi}
\author{L\'{a}szl\'{o} Szunyogh}
\affiliation{Department of Theoretical Physics, Budapest University of Technology and Economics, Budafoki \'{u}t 8, H-1111 Budapest, Hungary}
\affiliation{MTA-BME Condensed Matter Research Group, Budapest University of Technology and Economics, Budafoki \'{u}t 8, H-1111 Budapest, Hungary}
\author{Levente R\'{o}zsa}
\affiliation{University of Konstanz, Konstanz, Germany}

\date{\today}

\begin{abstract}
Controlling the preferred direction of the magnetic moments is essential for the design of spintronic devices based on ultrathin films and heterostructures. As the film thickness or the temperature is increased, the easy anisotropy axis is typically reoriented from an out-of-plane direction preferred by surface and interface energy contributions to an in-plane alignment favored by the volume anisotropy terms. We study the temperature-driven spin reorientation transition in two atomic layers of Fe on W(110) using well-tempered metadynamics simulations based on a spin model parametrized by \textit{ab initio} calculations and find that the transition only takes place in the presence of the Dzyaloshinsky--Moriya interaction (DMI). This demonstrates that the chiral DMI does not only differentiate between noncollinear spin structures of different rotational senses, but it also influences the magnetic orientation of collinear magnetic configurations.
\end{abstract}

\maketitle

\section{Introduction}
Ultrathin magnetic films, heterostructures and nano\-particles have an important role in the development of energy-efficient and ultrahigh-density memory and logic devices. The combination of magnetic materials with heavy nonmagnetic elements displaying strong spin--orbit coupling (SOC) enables the fast electrical manipulation and detection of spin signals, being explored in the emerging field of spin-orbitronics. The SOC gives rise to the magnetocrystalline anisotropy (MCA) energy stabilizing nanoscale magnetic domains, particularly in thin films with perpendicular magnetic anisotropy (PMA). The Dzyaloshinsky--Moriya interaction (DMI) is also attributed to the SOC in the case of broken inversion symmetry, which naturally occurs at magnetic-nonmagnetic interfaces. The DMI is responsible for the formation of chiral noncollinear spin structures including spin spirals \cite{Bode2007}, domain walls \cite{Ryu2013} and skyrmions \cite{Muhlbauer2009,Romming2013}, and leads to a nonreciprocal propagation of spin waves \cite{Zakeri}.

The MCA competes with the shape anisotropy (SA) due to the magnetostatic dipole-dipole interaction, the latter preferring an in-plane orientation of the spins in layered systems. 
The magnetization direction of thin ferromagnetic films with PMA rotates in-plane as the film thickness is increased, which effect is called a spin reorientation transition (SRT). Increasing the temperature may also induce a SRT due to the different temperature dependence of the MCA and the SA \cite{Jensen}. Since its discovery in the 1960s in NiFe(111)/Cu(111) films \cite{Gradmann1968}, thickness- and temperature-driven SRTs have been observed in a wide range of nanostructures \cite{Jensen,Wilgocka2010,Kukunin2007,Farle1999}. Understanding the SRT is essential in engineering the thickness of the layers in heterostructures for stable room-temperature spintronic applications.

Ultrathin Fe films deposited on W(110) represent an extensively studied model system for SOC phenomena at a magnetic-nonmagnetic metal interface. The Fe monolayer grows pseudomorphically on the W(110) surface and exhibits an in-plane MCA along the $(1\overline{1}0)$ direction. The magnetic ground state of the double layer (DL) strongly depends on the size and shape of the DL areas in the experiments \cite{Bergmann2006} and on the strain relief \cite{Sander1999,Seo2011}, and it is sensitive to gas adsorption \cite{Durkop1997}. Larger DL islands are perpendicularly magnetized along the $[110]$ direction at low temperature \cite{Weber1997,Bergmann2006,Seo2011} turning into the $[1\overline{1}0]$ in-plane direction at higher temperature following a SRT \cite{Dunlavy2004}. The preferred magnetization direction rotates toward the $[001]$ direction at higher film thicknesses \cite{Slezak2010}. The Fe DL is one of the first ultrathin film systems where \textit{ab initio} calculations and experimental evidence determined the presence of chiral domain walls \cite{Heide2008,Meckler2009} and nonreciprocal magnon propagation \cite{Zakeri,Udvardi}, both of which arise due to the interfacial DMI and have become established signatures for detecting the chiral magnetic interaction.

Although both the MCA and the DMI are attributed to the SOC, they typically compete with each other since the former tends to align all spins along a preferential direction, while the latter opens a finite angle between the magnetic moments. However, in inhomogeneous and disordered systems the DMI may also enhance the anisotropy field, such as in spin glasses containing nonmagnetic heavy metal impurities and displaying a noncollinear magnetic configuration \cite{Fert,Levy}, or due to the spin canting at the edges of nanomagnets \cite{Cubukcu}. In Ref.~\cite{Rozsa}, it has been shown that thermal fluctuations and spin correlations lead to the emergence of a DMI-induced anisotropy term even in homogeneous ferromagnetic systems. The interplay between the anisotropy and the DMI also lends a chiral character to the exchange bias effect~\cite{Yanes,Fernandez-Pacheco2019}, and it was recently observed to give rise to a modulated ordered phase at a temperature-driven SRT~\cite{Dashwood}.

In this work we demonstrate that the DMI may lead to a temperature-induced SRT in homogeneous collinear magnetic systems, by giving rise to an anisotropy term competing with the MCA and the SA. By performing well-tempered metadynamics simulations using an atomistic spin model parametrized by \textit{ab initio} calculations, we reproduce the experimentally observed SRT in a DL Fe on W(110), and show that it does not occur if the DMI is absent. These findings highlight the role of spin correlations in determining equilibrium magnetic configurations at finite temperature, which should be relevant for room-temperature applications based on chiral domain walls or skyrmions.

\section{Computational details}

\subsection{Spin model}

The DL Fe on W(110) is modeled by the classical spin Hamiltonian
\begin{eqnarray} \label{eq:HFe2W}
	   H &=&  \frac{1}{2}\sum_{p,q=1}^2\sum_{i\neq j} \mathbf{s}_{pi}^T\mathbf{J}_{pi,qj}\mathbf{s}_{qj}  \\
     & \ & + \sum_{p=1}^2 \sum_i\lambda_{px} (\mathbf{s}_{pi}\hat{\mathbf{x}})^2
      +  \sum_{p=1}^2 \sum_i\lambda_{py} (\mathbf{s}_{pi}\hat{\mathbf{y}})^2 \;, \nonumber 
 \end{eqnarray}
where $\mathbf{s}_{pi}$ is a unit vector representing the direction of the atomic magnetic moment at
site $i$ in layer $p$. $\mathbf{J}_{pi,qj}$ is a $3\times 3$ tensor of exchange interactions, which can be decomposed as \cite{Udvardi2003}
\begin{eqnarray}
\mathbf{J}_{pi,qj}&=&\frac{1}{3}\textrm{Tr}\left(\mathbf{J}_{pi,qj}\right)\mathbf{I}+\frac{1}{2}\left(\mathbf{J}_{pi,qj}-\mathbf{J}^{T}_{pi,qj}\right)\nonumber
\\
& \ &+\frac{1}{2}\left(\mathbf{J}_{pi,qj}+\mathbf{J}^{T}_{pi,qj}-\frac{2}{3}\textrm{Tr}\left(\mathbf{J}_{pi,qj}\right)\mathbf{I}\right),\label{Jij}
\end{eqnarray}
where $J_{pi,qj}=\frac{1}{3}\textrm{Tr}\left(\mathbf{J}_{pi,qj}\right)$ is the isotropic Heisenberg exchange coupling, $\mathbf{s}^{T}_{pi}\frac{1}{2}\left(\mathbf{J}_{pi,qj}-\mathbf{J}^{T}_{pi,qj}\right)\mathbf{s}_{qj}=\mathbf{D}_{pi,qj}\left(\mathbf{s}_{pi}\times\mathbf{s}_{qj}\right)$ is the antisymmetric DMI, and the third term, corresponding to a symmetric traceless matrix, stands for both the classical dipolar and the SOC-induced pseudo-dipolar interactions. The on-site biaxial MCA is characterized by the constants $\lambda_{px}$ and $\lambda_{py}$, where $\hat{\mathbf{x}}$ and $\hat{\mathbf{y}}$ are unit vectors parallel to the $[1\overline{1}0]$ and $[001]$ directions, respectively.


\subsection{\textit{Ab initio} calculations}

The electronic structure of the system was determined via the screened Korringa--Kohn--Rostoker (SKKR) method \cite{skkr}. The seven atomic layers of W, two atomic layers of Fe and three atomic layers of empty spheres (vacuum) were located between semi-infinite bulk W and semi-infinite vacuum. The bulk lattice constant of bcc Fe is $a_\mathrm{Fe}=2.867\,$\AA, and of bcc W is $a_\mathrm{W}=3.165\,$\AA. Due to this large lattice mismatch, there is a considerable inward relaxation of the Fe layers, which has been confirmed by experimental\cite{Albrecht1991,Tober1997,Meyerheim2001} as well as theoretical \cite{Qian1999,Santos2016,Heide2006} investigations. In our calculations the distance between the top W layer and the interfacial Fe layer was chosen to be $2.01\,$\AA, while the distance between the interfacial and surface Fe layers was set to $1.71\,$\AA, according to the results of earlier studies based on density functional theory calculations \cite{Santos2016,Heide2006}. 
The spin magnetic moments of the Fe layers were found to be $\mu_{\textrm{S}}=2.78\,\mu_{\textrm{B}}$ in the surface layer and $\mu_{\textrm{I}}=2.34\,\mu_{\textrm{B}}$ in the interface layer, and the ferromagnetic alignment of the layers was preferred. The largest induced magnetic moment of $\mu_{\textrm{W}}=0.16\,\mu_{\textrm{B}}$ was found in the top W layer, and it was aligned antiparallel to the magnetic layers.

The relativistic torque method \cite{Udvardi2003} was applied to calculate the exchange interaction tensors and the on-site magnetocrystalline anisotropy coefficients. The calculations were performed in parallel alignments of the spins along two in-plane nearest-neighbor directions, one in-plane next-nearest-neighbor and third-nearest-neighbor direction, and along the out-of-plane directions. The ferromagnetic or antiferromagnetic alignment of the moments in the Fe as well as the substrate layers was based on the self-consistent calculations. Since orienting the moments along a specific direction only gives information about the exchange tensor elements with Cartesian indices perpendicular to the magnetization direction, a least-squares fitting procedure was applied to the results of the five different calculations mentioned above. A real-space cut-off of $8\sqrt{2}a_\mathrm{W}=35.808\,$\AA\: was set as a maximal in-plane distance between interacting atoms. The magnetostatic dipolar interaction was added to the traceless symmetric part of the exchange tensors, since it is not taken into account in the \textit{ab initio} calculations. The dipolar term favors spin alignment along the $[001]$ direction by $0.143\,\textrm{meV/atom}$ and along the $[1\overline{1}0]$ direction by $0.137\,\textrm{meV/atom}$ over the out-of-plane $[\overline{1}\overline{1}0]$ orientation. Due to the finite cut-off of the interaction range, these anisotropy contributions are approximately $10\,\mu\textrm{eV/atom}$ lower at $T=0\,\textrm{K}$ compared to the value calculated for an infinite DL based on the Madelung constants; see, e.g., Ref.~\cite{skkr} for the calculation method.

The layer-resolved anisotropy parameters and the exchange tensors $\mathbf{J}_{pi,qj}$ are provided in the Supplemental File~\cite{supp}.

\subsection{Metadynamics simulations} 

The finite-temperature magnetic anisotropy energy was calculated using well-tempered metadynamics \cite{Laio2002, Barducci2008}. The computational scheme is described in detail in Ref.~\cite{Nagyfalusi2019}. During the simulations, the free energy is sampled along a collective variable $\eta = M_z/M$, where $M_z = \sum_{p,i} (\mathbf{s}_{pi}\hat{\mathbf{z}})$ is the $z$ component of the magnetization, $\mathbf{M}=\sum_{p,i} \mathbf{s}_{p,i}$ and $M = \vert \mathbf{M}\vert$. Following the notations used in Ref.~\cite{Nagyfalusi2019}, the height of the Gaussian bias potentials was $w_0=0.04$\,mRy, their width was $\sigma=0.06$, and the metadynamics temperature was set to $T_\mathrm{m}=10400\,$K. The simulations were performed on a $64\times64\times2$ lattice.

\section{Results}

The calculated on-site and two-site anisotropy parameters are listed in Table \ref{tbl1}. The two-site anisotropy parameters are defined as
\begin{equation}
 {J}^{\alpha\alpha} = \frac{1}{4}\sum_{p,q=1,2} \sum_j {J}^{\alpha\alpha}_{p0,qj} \;,
\end{equation}
where $\alpha=x,y,z$ is a Cartesian component and $0$ is a reference site in layer $p$. Note that ${J}^{\alpha\alpha}$ includes the contribution from the SA and the on-site coefficients are averaged over the two Fe layers. From these parameters the energy difference between the $[1\overline{1}0]$ and $[\overline{1}\overline{1}0]$ directions of the magnetization yields 0.170 meV/Fe atom, while between the $[001]$ and $[\overline{1}\overline{1}0]$ directions it is 0.189 meV/Fe atom. The ground state magnetic configuration is therefore out-of-plane, in agreement with the experimental observations \cite{Weber1997,Bergmann2006,Seo2011}. The competition between the on-site and two-site anisotropy terms, however, foreshadows a possible temperature-driven SRT.

\begin{table}
 \caption{\label{tbl1} Calculated magnetic anisotropy parameters per Fe atom for the DL Fe on W(110). The values are given in meV. The notations \textit{x}, \textit{y} and \textit{z} denote the $[1\overline{1}0]$, $[001]$ and $[\overline{1}\overline{1}0]$ directions, respectively. 
 }
\begin{ruledtabular}
 \begin{tabular}{cccc}
   $\lambda_x$                        & $\lambda_y$                       & $J^{xx} - J^{zz}$ & $J^{yy} - J^{zz}$ \\\hline
   0.283     & 0.067 &-0.113 &0.122  \\ 
 \end{tabular}
\end{ruledtabular}
\end{table}


\begin{figure}
   \includegraphics[width=\columnwidth]{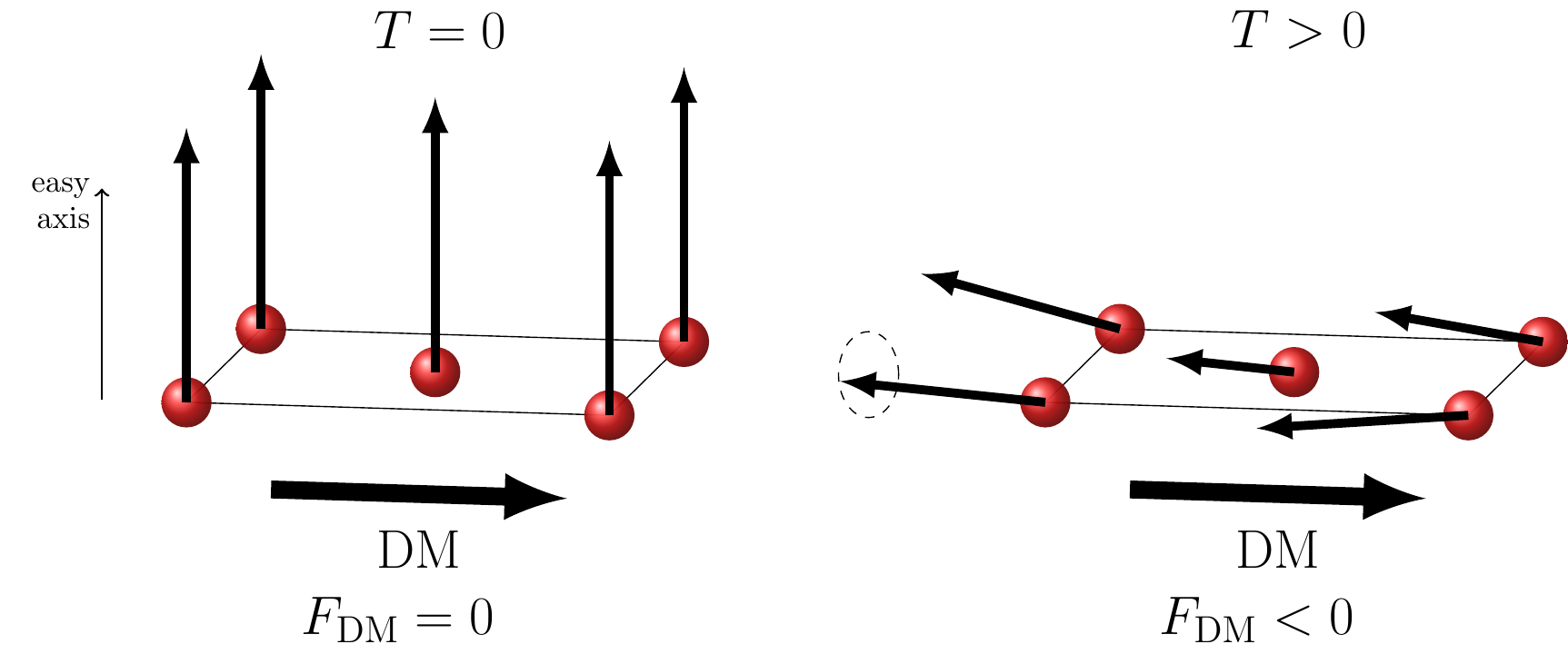}
   \caption{\label{fig1}SRT induced by the DMI in the DL Fe on W(110). The free energy of the system is minimized by chiral spin fluctuations in the plane perpendicular to the DM vectors. The DM vectors lie in the surface plane because of the $C_{2\textrm{v}}$ symmetry of the system.
 }
\end{figure} 

While the DMI does not contribute to the MAE of a collinear magnetic configuration at zero temperature, our finite-temperature simulations discussed below give clear evidence that the SRT is only facilitated by the presence of the DMI in this system. The mechanism behind the DMI-induced SRT is illustrated in Fig.~\ref{fig1}. At finite temperature, the spins minimize the free energy by fluctuating around their equilibrium directions. To estimate the free energy gain from the DMI, we use the formula
\begin{align}
K_{\textrm{DMI}} =  \frac{m}{2}\sum\limits_{p,q=1}^2\sum_{j}\mathbf{D}_{p0,qj}\cdot\left<\mathbf{s}_{p0}\times\mathbf{s}_{qj}\right>\, ,
 \label{eq:ss_corr}
\end{align}
where $m$ is the dimensionless magnetization, normalized to $m=1$ at zero temperature. 
The correlation function corresponding to the vector chirality $\left<\mathbf{s}_{p,i}\times\mathbf{s}_{q,j}\right>$ describes transversal fluctuations around the ferromagnetic state, which acquire a chiral character due to the DMI. Since the DMI energy contribution reads $\mathbf{D}_{pi,qj}\left(\mathbf{s}_{pi}\times\mathbf{s}_{qj}\right)$, the scalar product $\mathbf{D}_{pi,qj}\cdot\left<\mathbf{s}_{p,i}\times\mathbf{s}_{q,j}\right>$ will be negative when summed up over all neighbors in order to minimize the free energy. Free energy may be gained if the magnetization 
is oriented parallel to the DM vector, since the chiral fluctuations have to take place in a plane perpendicular to both the magnetization and the DM vector. Therefore, Eq.~\eqref{eq:ss_corr} depends on the orientation of the magnetization and may be interpreted as an anisotropy term preferring the alignment of the spins parallel to the DM vector, the direction of which is restricted by the symmetries of the system. The effect of this anisotropy term on the spin wave spectrum was discussed in Ref.~\cite{Rozsa}.

The components of the largest DM vectors in the DL Fe on W(110) determined from the \textit{ab initio} calculations are listed in Table~\ref{tbl2}. The $C_{2\textrm{v}}$ symmetry of the system causes the in-plane components of the DM vectors to dominate, as expected for interfacial inversion-symmetry breaking. The $z$ component of the DMI is exactly zero for all pairs in the same magnetic layer, as well as for interlayer pairs located in symmetry planes. Overall the $x$ component of the DM vectors has the largest contribution, preferring the formation of right-handed N\'{e}el-type domain walls with normal vectors along the $[001]$ direction in the out-of-plane magnetized system, in agreement with the experimental observations \cite{Meckler2009}.
  
  \begin{table}
 \caption{\label{tbl2} In-plane components of the DM vectors with the largest magnitude in the DL Fe on W(110). The Fe layer at the interface with W is denoted by I, the Fe at the surface by S. The number after N denotes the order of the neighbor. $D^x_{pi,qj}>0$ prefers a right-handed rotation or fluctuation of the spins in the $yz$ plane ($z\rightarrow y$), $D^y_{pi,qj}>0$ describes an energy gain from a left-handed rotation in the $xz$ plane ($x\rightarrow z$), following the convention of Ref.~\cite{Meckler2009}. The $z$ component of the DM vectors is $0$ for all pairs in the table.}
\begin{ruledtabular}
 \begin{tabular}{ccc} 
  pair &  $D^x_{pi,qj}$ (meV)  & $D^y_{pi,qj}$ (meV)\\\hline
   S--S N1 &  1.62   & -1.98  \\
   S--I N1 &  1.69   & 0.00  \\
   I--I N2 & 1.22    & 0.00  \\
   I--I N4 & -1.12   &  0.34  \\
 \end{tabular}
\end{ruledtabular}
\end{table}

Regarding the values in Table~\ref{tbl2}, Eq.~\eqref{eq:ss_corr} indicates that at finite temperatures where the spin fluctuations play a prominent role 
the DMI prefers the 
magnetization to be aligned 
in-plane along the $x$ direction. This contributes to the competition between the on-site and the two-site anisotropy coefficients (cf. Table~\ref{tbl1}), and leads to the SRT which was observed experimentally \cite{Bergmann2006}.

\begin{figure}
   \includegraphics[width=\columnwidth]{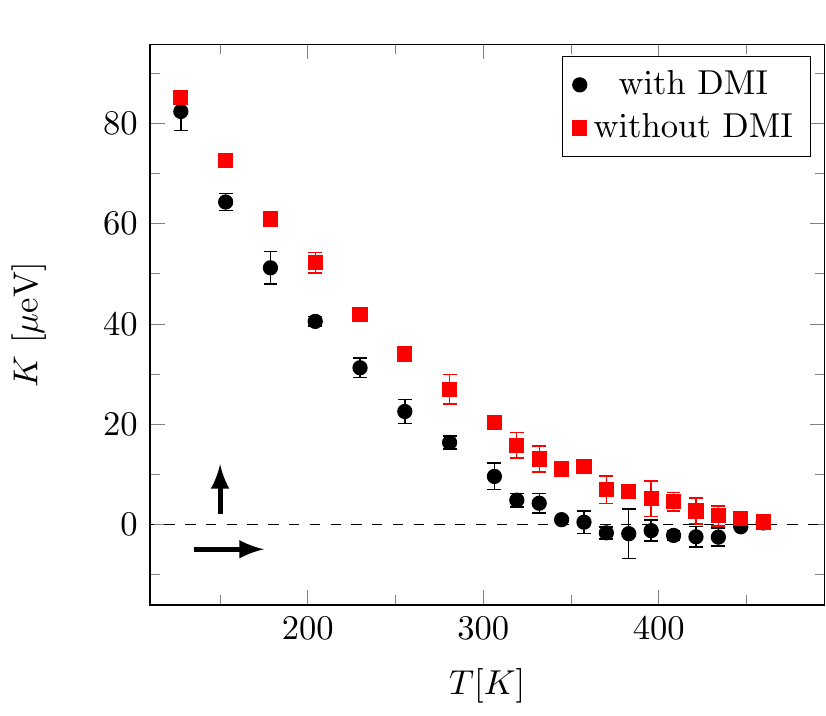}
   
   \caption{\label{fig2} 
   Temperature-dependent MAE of the DL Fe on W(110) obtained from metadynamics simulations, defined  as the difference in the free energy between the normal-to-plane and the in-plane magnetic configurations. The presented values are normalized to one Fe atom. The black filled circles show results based on the full Hamiltonian in Eq.~\eqref{eq:HFe2W}, while the red squares correspond to a simulation where the DMI was set to zero. An average over 60 Monte Carlo simulations was performed at each temperature, the error bars denote the standard deviation calculated from these independent simulations. The number of Monte Carlo steps was at least $1.5\times10^{5}$ at each temperature, with a higher number of steps close to the phase transitions. The calculations were carried out on a $64\times64\times2$ lattice, using the metadynamics parameters $T_\mathrm{m}=10400\,\mathrm{K}$, $w_0=0.04\,\mathrm{mRy}$, and $\sigma=0.06$.}
 
\end{figure}

The presence of the SRT in the model described by Eq.~\eqref{eq:HFe2W} was confirmed by calculating the temperature-dependent magnetic anisotropy energy (MAE) $K$, defined as the free-energy difference per spin between in-plane and normal-to-plane magnetic configurations. This was performed via well-tempered metadynamics simulations \cite{Laio2002, Barducci2008, Nagyfalusi2019}, a Monte Carlo-based method enabling to sample the free energy surface. As shown in Fig.~\ref{fig2}, the anisotropy parameter $K$ decreases with temperature and changes sign around $T_{\textrm{r}}\approx 350\,$K, indicating a SRT transition from the normal-to-plane ($K>0$) to the in-plane ($K<0$) direction. The temperature of the SRT is higher than the value reported experimentally in Ref.~\cite{Bergmann2006} ($T_{\textrm{r}}\approx 200\,$K), but is supported by the presence of out-of-plane magnetized DL patches found at room temperature in Ref.~\cite{Slezak2013}, demonstrating that $T_{\textrm{r}}$ strongly depends on the size and shape of the DL areas. Although the metadynamics simulations which use only the out-of-plane magnetization component as a collective parameter do not enable to differentiate between in-plane directions, calculations based on the Metropolis Monte Carlo algorithm confirmed that above $T_{\textrm{r}}$ the average magnetization is aligned along the $x$ direction, in agreement with the prediction based on Table~\ref{tbl1}.  The anisotropy goes to zero around the Curie temperature $T_{\textrm{C}}\approx 450\,$K, above which a paramagnetic phase is found in the Metropolis Monte Carlo simulations. The simulated critical temperature is in good quantitative agreement with the measurements in Ref.~\cite{Dunlavy2004}.

In order to quantify the role of the DMI in the SRT, the simulations were repeated after removing the antisymmetric part of the coupling tensors $\mathbf{J}_{pi,qj}$. As demonstrated in Fig.~\ref{fig2}, in this case the MAE is 
shifted upwards 
at all temperatures compared to simulations utilizing the full coupling, as expected from Eq.~\eqref{eq:ss_corr} since the in-plane DM vectors prefer an in-plane orientation of the spins. Importantly, the MAE does not change sign in the simulations without DMI. This confirms that the SRT in the Fe DL on W(110) can only be explained if the contribution of the DMI to the anisotropy energy at finite temperature is taken into account.

\begin{figure}
  \includegraphics[width=\columnwidth]{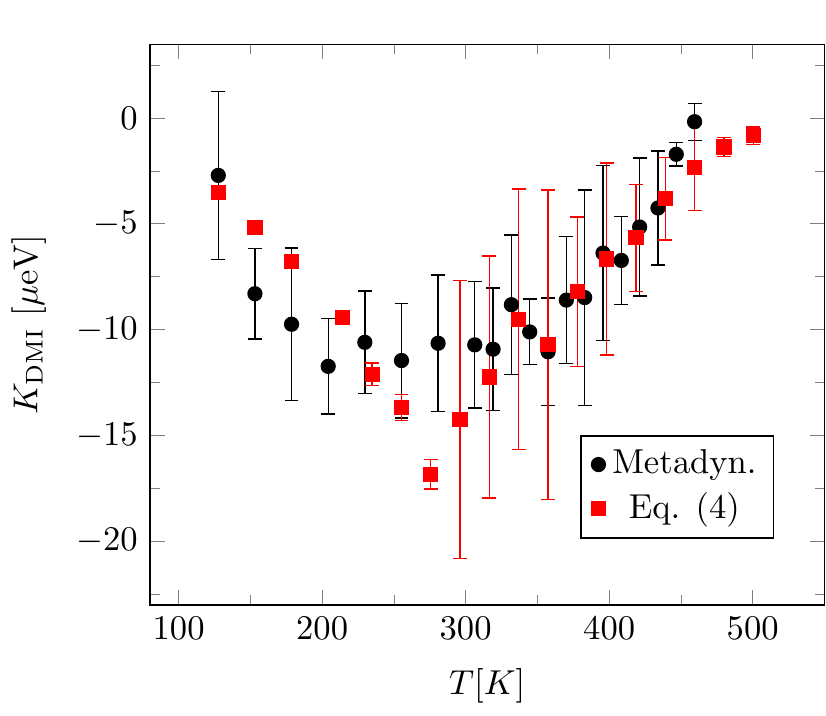}
 \caption{\label{fig3}  The difference in the MAE per Fe atom between metadynamics simulations with and without the DMI (black filled circles). The red squares denote $K_{\textrm{DMI}}$ from Eq.~\eqref{eq:ss_corr}, which attributes this anisotropy to the DMI-induced chiral spin correlations, with the magnetization and the correlation function determined from Metropolis Monte Carlo simulations. Both calculations were performed on a $64\times64\times2$ lattice. Error bars denote the standard deviation calculated from 60 independent simulations.}
\end{figure}

The MAE difference for the simulations performed with and without DMI is displayed in Fig.~\ref{fig3}, along with the theoretical prediction of $K_{\textrm{DMI}}$ from Eq.~\eqref{eq:ss_corr}. The magnetization and the correlation functions necessary for the calculation of $K_{\textrm{DMI}}$ were obtained from Metropolis Monte Carlo simulations, with the correlation function determined only for the neighbors with the largest DMI interactions listed in Table~\ref{tbl2}. Figure~\ref{fig3} demonstrates that the analytical expression for $K_{\textrm{DMI}}$, which only takes into account the lowest-order correlation corrections, is in good quantitative agreement with the full-scale numerical simulations of the anisotropy contribution attributed to the DMI over the whole temperature range. Importantly, this contribution approaches zero at low temperature where the fluctuations are suppressed, and it also decreases close to the Curie temperature where the magnetization disappears, as can be deduced from Eq.~\eqref{eq:ss_corr}. The DMI-induced anisotropy is maximal around room temperature in the system, close to the SRT temperature. This is fundamentally different from the temperature dependence of the on-site MCA and the SA, which monotonically decrease from zero temperature, and for which the spin--spin correlations lead to a faster decay compared to the mean-field prediction \cite{Evans2020}.

\section{Conclusions}
In summary, we demonstrated that taking into account the chiral spin correlations induced by the DMI is necessary to explain the experimentally observed temperature-induced SRT in the DL Fe on W(110). These spin correlations induce an additional anisotropy term preferring to align the spins along the directions of the DM vectors, mostly lying in the plane of the sample in ultrathin films and multilayer systems with inversion-symmetry breaking at the interface. Characteristically, this DMI-induced anisotropy term increases with temperature and reaches its maximum value close to room temperature in the considered system, while other anisotropy terms monotonically decrease in magnitude as the temperature becomes higher, and the correlation corrections enhance this decay compared to the mean-field predictions \cite{Evans2020}. Therefore, it is expected that this DMI-induced anisotropy term should have a considerable influence on the design of room-temperature spintronic applications, by affecting the magnon frequencies, the width of domain walls and the sizes of magnetic skyrmions. The possible competition between intralayer and interlayer contributions to the DMI \cite{Han2019,Fernandez-Pacheco2019,Vedmedenko2019} should further enrich the variety of observed phenomena in sputtered multilayers, where the growth process typically breaks additional symmetries compared to epitaxial ultrathin films.
 
 \begin{acknowledgments}
The authors are grateful for the financial support of the National Research, Development, and Innovation Office of Hungary (NKFIH) under Project No.  K131938, the BME Nanotechnology and Materials Science TKP2020 IE grant of NKFIH Hungary  (BME IE-NAT TKP2020), and of the Alexander von Humboldt Foundation.
\end{acknowledgments}


\begin{thebibliography}{43}%
\makeatletter
\providecommand \@ifxundefined [1]{%
 \@ifx{#1\undefined}
}%
\providecommand \@ifnum [1]{%
 \ifnum #1\expandafter \@firstoftwo
 \else \expandafter \@secondoftwo
 \fi
}%
\providecommand \@ifx [1]{%
 \ifx #1\expandafter \@firstoftwo
 \else \expandafter \@secondoftwo
 \fi
}%
\providecommand \natexlab [1]{#1}%
\providecommand \enquote  [1]{``#1''}%
\providecommand \bibnamefont  [1]{#1}%
\providecommand \bibfnamefont [1]{#1}%
\providecommand \citenamefont [1]{#1}%
\providecommand \href@noop [0]{\@secondoftwo}%
\providecommand \href [0]{\begingroup \@sanitize@url \@href}%
\providecommand \@href[1]{\@@startlink{#1}\@@href}%
\providecommand \@@href[1]{\endgroup#1\@@endlink}%
\providecommand \@sanitize@url [0]{\catcode `\\12\catcode `\$12\catcode
  `\&12\catcode `\#12\catcode `\^12\catcode `\_12\catcode `\%12\relax}%
\providecommand \@@startlink[1]{}%
\providecommand \@@endlink[0]{}%
\providecommand \url  [0]{\begingroup\@sanitize@url \@url }%
\providecommand \@url [1]{\endgroup\@href {#1}{\urlprefix }}%
\providecommand \urlprefix  [0]{URL }%
\providecommand \Eprint [0]{\href }%
\providecommand \doibase [0]{https://doi.org/}%
\providecommand \selectlanguage [0]{\@gobble}%
\providecommand \bibinfo  [0]{\@secondoftwo}%
\providecommand \bibfield  [0]{\@secondoftwo}%
\providecommand \translation [1]{[#1]}%
\providecommand \BibitemOpen [0]{}%
\providecommand \bibitemStop [0]{}%
\providecommand \bibitemNoStop [0]{.\EOS\space}%
\providecommand \EOS [0]{\spacefactor3000\relax}%
\providecommand \BibitemShut  [1]{\csname bibitem#1\endcsname}%
\let\auto@bib@innerbib\@empty
\bibitem [{\citenamefont {Bode}\ \emph {et~al.}(2007)\citenamefont {Bode},
  \citenamefont {Heide}, \citenamefont {von Bergmann}, \citenamefont
  {Ferriani}, \citenamefont {Heinze}, \citenamefont {Bihlmayer}, \citenamefont
  {Kubetzka}, \citenamefont {Pietzsch}, \citenamefont {Bl\"{u}gel},\ and\
  \citenamefont {Wiesendanger}}]{Bode2007}%
  \BibitemOpen
  \bibfield  {author} {\bibinfo {author} {\bibfnamefont {M.}~\bibnamefont
  {Bode}}, \bibinfo {author} {\bibfnamefont {M.}~\bibnamefont {Heide}},
  \bibinfo {author} {\bibfnamefont {K.}~\bibnamefont {von Bergmann}}, \bibinfo
  {author} {\bibfnamefont {P.}~\bibnamefont {Ferriani}}, \bibinfo {author}
  {\bibfnamefont {S.}~\bibnamefont {Heinze}}, \bibinfo {author} {\bibfnamefont
  {G.}~\bibnamefont {Bihlmayer}}, \bibinfo {author} {\bibfnamefont
  {A.}~\bibnamefont {Kubetzka}}, \bibinfo {author} {\bibfnamefont
  {O.}~\bibnamefont {Pietzsch}}, \bibinfo {author} {\bibfnamefont
  {S.}~\bibnamefont {Bl\"{u}gel}},\ and\ \bibinfo {author} {\bibfnamefont
  {R.}~\bibnamefont {Wiesendanger}},\ }\bibfield  {title} {\bibinfo {title}
  {Chiral magnetic order at surfaces driven by inversion asymmetry},\ }\href
  {https://doi.org/10.1038/nature05802} {\bibfield  {journal} {\bibinfo
  {journal} {Nature}\ }\textbf {\bibinfo {volume} {447}},\ \bibinfo {pages}
  {190} (\bibinfo {year} {2007})}\BibitemShut {NoStop}%
\bibitem [{\citenamefont {Ryu}\ \emph {et~al.}(2013)\citenamefont {Ryu},
  \citenamefont {Thomas}, \citenamefont {Yang},\ and\ \citenamefont
  {Parkin}}]{Ryu2013}%
  \BibitemOpen
  \bibfield  {author} {\bibinfo {author} {\bibfnamefont {K.-S.}\ \bibnamefont
  {Ryu}}, \bibinfo {author} {\bibfnamefont {L.}~\bibnamefont {Thomas}},
  \bibinfo {author} {\bibfnamefont {S.-H.}\ \bibnamefont {Yang}},\ and\
  \bibinfo {author} {\bibfnamefont {S.}~\bibnamefont {Parkin}},\ }\bibfield
  {title} {\bibinfo {title} {Chiral spin torque at magnetic domain walls},\
  }\href {https://doi.org/10.1038/nnano.2013.102} {\bibfield  {journal}
  {\bibinfo  {journal} {Nat. Nanotechnol.}\ }\textbf {\bibinfo {volume} {8}},\
  \bibinfo {pages} {527} (\bibinfo {year} {2013})}\BibitemShut {NoStop}%
\bibitem [{\citenamefont {M\"{u}hlbauer}\ \emph {et~al.}(2009)\citenamefont
  {M\"{u}hlbauer}, \citenamefont {Binz}, \citenamefont {Jonietz}, \citenamefont
  {Pfleiderer}, \citenamefont {Rosch}, \citenamefont {Neubauer}, \citenamefont
  {Georgii},\ and\ \citenamefont {B\"{o}ni}}]{Muhlbauer2009}%
  \BibitemOpen
  \bibfield  {author} {\bibinfo {author} {\bibfnamefont {S.}~\bibnamefont
  {M\"{u}hlbauer}}, \bibinfo {author} {\bibfnamefont {B.}~\bibnamefont {Binz}},
  \bibinfo {author} {\bibfnamefont {F.}~\bibnamefont {Jonietz}}, \bibinfo
  {author} {\bibfnamefont {C.}~\bibnamefont {Pfleiderer}}, \bibinfo {author}
  {\bibfnamefont {A.}~\bibnamefont {Rosch}}, \bibinfo {author} {\bibfnamefont
  {A.}~\bibnamefont {Neubauer}}, \bibinfo {author} {\bibfnamefont
  {R.}~\bibnamefont {Georgii}},\ and\ \bibinfo {author} {\bibfnamefont
  {P.}~\bibnamefont {B\"{o}ni}},\ }\bibfield  {title} {\bibinfo {title}
  {Skyrmion lattice in a chiral magnet},\ }\href
  {https://doi.org/10.1126/science.1166767} {\bibfield  {journal} {\bibinfo
  {journal} {Science}\ }\textbf {\bibinfo {volume} {323}},\ \bibinfo {pages}
  {915} (\bibinfo {year} {2009})}\BibitemShut {NoStop}%
\bibitem [{\citenamefont {Romming}\ \emph {et~al.}(2013)\citenamefont
  {Romming}, \citenamefont {Hanneken}, \citenamefont {Menzel}, \citenamefont
  {Bickel}, \citenamefont {Wolter}, \citenamefont {von Bergmann}, \citenamefont
  {Kubetzka},\ and\ \citenamefont {Wiesendanger}}]{Romming2013}%
  \BibitemOpen
  \bibfield  {author} {\bibinfo {author} {\bibfnamefont {N.}~\bibnamefont
  {Romming}}, \bibinfo {author} {\bibfnamefont {C.}~\bibnamefont {Hanneken}},
  \bibinfo {author} {\bibfnamefont {M.}~\bibnamefont {Menzel}}, \bibinfo
  {author} {\bibfnamefont {J.~E.}\ \bibnamefont {Bickel}}, \bibinfo {author}
  {\bibfnamefont {B.}~\bibnamefont {Wolter}}, \bibinfo {author} {\bibfnamefont
  {K.}~\bibnamefont {von Bergmann}}, \bibinfo {author} {\bibfnamefont
  {A.}~\bibnamefont {Kubetzka}},\ and\ \bibinfo {author} {\bibfnamefont
  {R.}~\bibnamefont {Wiesendanger}},\ }\bibfield  {title} {\bibinfo {title}
  {Writing and deleting single magnetic skyrmions},\ }\href
  {https://doi.org/10.1126/science.1240573} {\bibfield  {journal} {\bibinfo
  {journal} {Science}\ }\textbf {\bibinfo {volume} {341}},\ \bibinfo {pages}
  {636} (\bibinfo {year} {2013})}\BibitemShut {NoStop}%
\bibitem [{\citenamefont {Zakeri}\ \emph {et~al.}(2010)\citenamefont {Zakeri},
  \citenamefont {Zhang}, \citenamefont {Prokop}, \citenamefont {Chuang},
  \citenamefont {Sakr}, \citenamefont {Tang},\ and\ \citenamefont
  {Kirschner}}]{Zakeri}%
  \BibitemOpen
  \bibfield  {author} {\bibinfo {author} {\bibfnamefont {K.}~\bibnamefont
  {Zakeri}}, \bibinfo {author} {\bibfnamefont {Y.}~\bibnamefont {Zhang}},
  \bibinfo {author} {\bibfnamefont {J.}~\bibnamefont {Prokop}}, \bibinfo
  {author} {\bibfnamefont {T.-H.}\ \bibnamefont {Chuang}}, \bibinfo {author}
  {\bibfnamefont {N.}~\bibnamefont {Sakr}}, \bibinfo {author} {\bibfnamefont
  {W.~X.}\ \bibnamefont {Tang}},\ and\ \bibinfo {author} {\bibfnamefont
  {J.}~\bibnamefont {Kirschner}},\ }\bibfield  {title} {\bibinfo {title}
  {Asymmetric spin-wave dispersion on Fe(110): Direct evidence of the
  dzyaloshinskii-moriya interaction},\ }\href
  {https://doi.org/10.1103/PhysRevLett.104.137203} {\bibfield  {journal}
  {\bibinfo  {journal} {Phys. Rev. Lett.}\ }\textbf {\bibinfo {volume} {104}},\
  \bibinfo {pages} {137203} (\bibinfo {year} {2010})}\BibitemShut {NoStop}%
\bibitem [{\citenamefont {Jensen}\ and\ \citenamefont
  {Bennemann}(2006)}]{Jensen}%
  \BibitemOpen
  \bibfield  {author} {\bibinfo {author} {\bibfnamefont {P.}~\bibnamefont
  {Jensen}}\ and\ \bibinfo {author} {\bibfnamefont {K.}~\bibnamefont
  {Bennemann}},\ }\bibfield  {title} {\bibinfo {title} {Magnetic structure of
  films: Dependence on anisotropy and atomic morphology},\ }\href
  {https://doi.org/https://doi.org/10.1016/j.surfrep.2006.02.001} {\bibfield
  {journal} {\bibinfo  {journal} {Surf. Sci. Rep.}\ }\textbf {\bibinfo {volume}
  {61}},\ \bibinfo {pages} {129 } (\bibinfo {year} {2006})}\BibitemShut
  {NoStop}%
\bibitem [{\citenamefont {Gradmann}\ and\ \citenamefont
  {M\"{u}ller}(1968)}]{Gradmann1968}%
  \BibitemOpen
  \bibfield  {author} {\bibinfo {author} {\bibfnamefont {U.}~\bibnamefont
  {Gradmann}}\ and\ \bibinfo {author} {\bibfnamefont {J.}~\bibnamefont
  {M\"{u}ller}},\ }\bibfield  {title} {\bibinfo {title} {Flat ferromagnetic,
  epitaxial 48Ni/52Fe(111) films of few atomic layers},\ }\href
  {https://doi.org/10.1002/pssb.19680270133} {\bibfield  {journal} {\bibinfo
  {journal} {Phys. Stat. Sol. B}\ }\textbf {\bibinfo {volume} {27}},\ \bibinfo
  {pages} {313} (\bibinfo {year} {1968})}\BibitemShut {NoStop}%
\bibitem [{\citenamefont {Wilgocka-Slezak}\ \emph {et~al.}(2010)\citenamefont
  {Wilgocka-Slezak}, \citenamefont {Freindl}, \citenamefont {Kozio\l{}},
  \citenamefont {Matlak}, \citenamefont {Rams}, \citenamefont {Spiridis},
  \citenamefont {Slezak}, \citenamefont {Slezak}, \citenamefont {Zajac},\ and\
  \citenamefont {Korecki}}]{Wilgocka2010}%
  \BibitemOpen
  \bibfield  {author} {\bibinfo {author} {\bibfnamefont {D.}~\bibnamefont
  {Wilgocka-Slezak}}, \bibinfo {author} {\bibfnamefont {K.}~\bibnamefont
  {Freindl}}, \bibinfo {author} {\bibfnamefont {A.}~\bibnamefont {Kozio\l{}}},
  \bibinfo {author} {\bibfnamefont {K.}~\bibnamefont {Matlak}}, \bibinfo
  {author} {\bibfnamefont {M.}~\bibnamefont {Rams}}, \bibinfo {author}
  {\bibfnamefont {N.}~\bibnamefont {Spiridis}}, \bibinfo {author}
  {\bibfnamefont {M.}~\bibnamefont {Slezak}}, \bibinfo {author} {\bibfnamefont
  {T.}~\bibnamefont {Slezak}}, \bibinfo {author} {\bibfnamefont
  {M.}~\bibnamefont {Zajac}},\ and\ \bibinfo {author} {\bibfnamefont
  {J.}~\bibnamefont {Korecki}},\ }\bibfield  {title} {\bibinfo {title}
  {Thickness-driven polar spin reorientation transition in ultrathin Fe/Au(001)
  films},\ }\href {https://doi.org/10.1103/PhysRevB.81.064421} {\bibfield
  {journal} {\bibinfo  {journal} {Phys. Rev. B}\ }\textbf {\bibinfo {volume}
  {81}},\ \bibinfo {pages} {064421} (\bibinfo {year} {2010})}\BibitemShut
  {NoStop}%
\bibitem [{\citenamefont {Kukunin}\ \emph {et~al.}(2007)\citenamefont
  {Kukunin}, \citenamefont {Prokop},\ and\ \citenamefont
  {Elmers}}]{Kukunin2007}%
  \BibitemOpen
  \bibfield  {author} {\bibinfo {author} {\bibfnamefont {A.}~\bibnamefont
  {Kukunin}}, \bibinfo {author} {\bibfnamefont {J.}~\bibnamefont {Prokop}},\
  and\ \bibinfo {author} {\bibfnamefont {H.~J.}\ \bibnamefont {Elmers}},\
  }\bibfield  {title} {\bibinfo {title} {Temperature-driven spin reorientation
  transition in Fe/Mo(110) nanostructures},\ }\href
  {https://doi.org/10.1103/PhysRevB.76.134414} {\bibfield  {journal} {\bibinfo
  {journal} {Phys. Rev. B}\ }\textbf {\bibinfo {volume} {76}},\ \bibinfo
  {pages} {134414} (\bibinfo {year} {2007})}\BibitemShut {NoStop}%
\bibitem [{\citenamefont {Farle}\ \emph {et~al.}(1999)\citenamefont {Farle},
  \citenamefont {Platow}, \citenamefont {Kosubek},\ and\ \citenamefont
  {Baberschke}}]{Farle1999}%
  \BibitemOpen
  \bibfield  {author} {\bibinfo {author} {\bibfnamefont {M.}~\bibnamefont
  {Farle}}, \bibinfo {author} {\bibfnamefont {W.}~\bibnamefont {Platow}},
  \bibinfo {author} {\bibfnamefont {E.}~\bibnamefont {Kosubek}},\ and\ \bibinfo
  {author} {\bibfnamefont {K.}~\bibnamefont {Baberschke}},\ }\bibfield  {title}
  {\bibinfo {title} {Magnetic anisotropy of Co/Cu(111) ultrathin films},\
  }\href {https://doi.org/https://doi.org/10.1016/S0039-6028(99)00762-1}
  {\bibfield  {journal} {\bibinfo  {journal} {Surf. Sci.}\ }\textbf {\bibinfo
  {volume} {439}},\ \bibinfo {pages} {146 } (\bibinfo {year}
  {1999})}\BibitemShut {NoStop}%
\bibitem [{\citenamefont {von Bergmann}\ \emph {et~al.}(2006)\citenamefont {von
  Bergmann}, \citenamefont {Bode},\ and\ \citenamefont
  {Wiesendanger}}]{Bergmann2006}%
  \BibitemOpen
  \bibfield  {author} {\bibinfo {author} {\bibfnamefont {K.}~\bibnamefont {von
  Bergmann}}, \bibinfo {author} {\bibfnamefont {M.}~\bibnamefont {Bode}},\ and\
  \bibinfo {author} {\bibfnamefont {R.}~\bibnamefont {Wiesendanger}},\
  }\bibfield  {title} {\bibinfo {title} {Coverage-dependent spin reorientation
  transition temperature of the Fe double-layer on W(110) observed by scanning
  tunneling microscopy},\ }\href
  {https://doi.org/https://doi.org/10.1016/j.jmmm.2005.12.015} {\bibfield
  {journal} {\bibinfo  {journal} {J. Magn. Magn. Mater.}\ }\textbf {\bibinfo
  {volume} {305}},\ \bibinfo {pages} {279 } (\bibinfo {year}
  {2006})}\BibitemShut {NoStop}%
\bibitem [{\citenamefont {Sander}(1999)}]{Sander1999}%
  \BibitemOpen
  \bibfield  {author} {\bibinfo {author} {\bibfnamefont {D.}~\bibnamefont
  {Sander}},\ }\bibfield  {title} {\bibinfo {title} {The correlation between
  mechanical stress and magnetic anisotropy in ultrathin films},\ }\href
  {http://stacks.iop.org/0034-4885/62/i=5/a=204} {\bibfield  {journal}
  {\bibinfo  {journal} {Rep. Prog. Phys.}\ }\textbf {\bibinfo {volume} {62}},\
  \bibinfo {pages} {809} (\bibinfo {year} {1999})}\BibitemShut {NoStop}%
\bibitem [{\citenamefont {Seo}\ \emph {et~al.}(2011)\citenamefont {Seo},
  \citenamefont {Oh}, \citenamefont {Kim},\ and\ \citenamefont
  {Kuk}}]{Seo2011}%
  \BibitemOpen
  \bibfield  {author} {\bibinfo {author} {\bibfnamefont {J.}~\bibnamefont
  {Seo}}, \bibinfo {author} {\bibfnamefont {Y.}~\bibnamefont {Oh}}, \bibinfo
  {author} {\bibfnamefont {T.-H.}\ \bibnamefont {Kim}},\ and\ \bibinfo {author}
  {\bibfnamefont {Y.}~\bibnamefont {Kuk}},\ }\bibfield  {title} {\bibinfo
  {title} {Strain relaxation induced spin reorientation in fe films on
  w(110)},\ }\href {https://doi.org/10.1063/1.3657138} {\bibfield  {journal}
  {\bibinfo  {journal} {Appl. Phys. Lett.}\ }\textbf {\bibinfo {volume} {99}},\
  \bibinfo {pages} {182501} (\bibinfo {year} {2011})}\BibitemShut {NoStop}%
\bibitem [{\citenamefont {D\"{u}rkop}\ \emph {et~al.}(1997)\citenamefont
  {D\"{u}rkop}, \citenamefont {Elmers},\ and\ \citenamefont
  {Gradmann}}]{Durkop1997}%
  \BibitemOpen
  \bibfield  {author} {\bibinfo {author} {\bibfnamefont {T.}~\bibnamefont
  {D\"{u}rkop}}, \bibinfo {author} {\bibfnamefont {H.}~\bibnamefont {Elmers}},\
  and\ \bibinfo {author} {\bibfnamefont {U.}~\bibnamefont {Gradmann}},\
  }\bibfield  {title} {\bibinfo {title} {Adsorption-driven spin reorientation
  transition in sesquilayers of Fe(110) on W(110)},\ }\href
  {https://doi.org/https://doi.org/10.1016/S0304-8853(97)00114-5} {\bibfield
  {journal} {\bibinfo  {journal} {J. Magn. Magn. Mater.}\ }\textbf {\bibinfo
  {volume} {172}},\ \bibinfo {pages} {L1 } (\bibinfo {year}
  {1997})}\BibitemShut {NoStop}%
\bibitem [{\citenamefont {Weber}\ \emph {et~al.}(1997)\citenamefont {Weber},
  \citenamefont {Wagner}, \citenamefont {Elmers}, \citenamefont {Hauschild},\
  and\ \citenamefont {Gradmann}}]{Weber1997}%
  \BibitemOpen
  \bibfield  {author} {\bibinfo {author} {\bibfnamefont {N.}~\bibnamefont
  {Weber}}, \bibinfo {author} {\bibfnamefont {K.}~\bibnamefont {Wagner}},
  \bibinfo {author} {\bibfnamefont {H.~J.}\ \bibnamefont {Elmers}}, \bibinfo
  {author} {\bibfnamefont {J.}~\bibnamefont {Hauschild}},\ and\ \bibinfo
  {author} {\bibfnamefont {U.}~\bibnamefont {Gradmann}},\ }\bibfield  {title}
  {\bibinfo {title} {Nanoscale spatial switching of magnetic anisotropy in
  pseudomorphic Fe(110) on W(110)},\ }\href
  {https://doi.org/10.1103/PhysRevB.55.14121} {\bibfield  {journal} {\bibinfo
  {journal} {Phys. Rev. B}\ }\textbf {\bibinfo {volume} {55}},\ \bibinfo
  {pages} {14121} (\bibinfo {year} {1997})}\BibitemShut {NoStop}%
\bibitem [{\citenamefont {Dunlavy}\ and\ \citenamefont
  {Venus}(2004)}]{Dunlavy2004}%
  \BibitemOpen
  \bibfield  {author} {\bibinfo {author} {\bibfnamefont {M.~J.}\ \bibnamefont
  {Dunlavy}}\ and\ \bibinfo {author} {\bibfnamefont {D.}~\bibnamefont
  {Venus}},\ }\bibfield  {title} {\bibinfo {title} {Critical susceptibility
  exponent measured from Fe/W(110) bilayers},\ }\href
  {https://doi.org/10.1103/PhysRevB.69.094411} {\bibfield  {journal} {\bibinfo
  {journal} {Phys. Rev. B}\ }\textbf {\bibinfo {volume} {69}},\ \bibinfo
  {pages} {094411} (\bibinfo {year} {2004})}\BibitemShut {NoStop}%
\bibitem [{\citenamefont {\ifmmode \acute{S}\else
  \'{S}\fi{}l\ifmmode~\mbox{\k{e}}\else \k{e}\fi{}zak}\ \emph
  {et~al.}(2010)\citenamefont {\ifmmode \acute{S}\else
  \'{S}\fi{}l\ifmmode~\mbox{\k{e}}\else \k{e}\fi{}zak}, \citenamefont {\ifmmode
  \acute{S}\else \'{S}\fi{}l\ifmmode~\mbox{\k{e}}\else \k{e}\fi{}zak},
  \citenamefont {Zaj\k{a}c}, \citenamefont {Freindl}, \citenamefont
  {Kozio\l{}-Rachwa\l{}}, \citenamefont {Matlak}, \citenamefont {Spiridis},
  \citenamefont {Wilgocka-\ifmmode \acute{S}\else
  \'{S}\fi{}l\ifmmode~\mbox{\k{e}}\else \k{e}\fi{}zak}, \citenamefont
  {Partyka-Jankowska}, \citenamefont {Rennhofer}, \citenamefont {Chumakov},
  \citenamefont {Stankov}, \citenamefont {R\"uffer},\ and\ \citenamefont
  {Korecki}}]{Slezak2010}%
  \BibitemOpen
  \bibfield  {author} {\bibinfo {author} {\bibfnamefont {T.}~\bibnamefont
  {\ifmmode \acute{S}\else \'{S}\fi{}l\ifmmode~\mbox{\k{e}}\else
  \k{e}\fi{}zak}}, \bibinfo {author} {\bibfnamefont {M.}~\bibnamefont {\ifmmode
  \acute{S}\else \'{S}\fi{}l\ifmmode~\mbox{\k{e}}\else \k{e}\fi{}zak}},
  \bibinfo {author} {\bibfnamefont {M.}~\bibnamefont {Zaj\k{a}c}}, \bibinfo
  {author} {\bibfnamefont {K.}~\bibnamefont {Freindl}}, \bibinfo {author}
  {\bibfnamefont {A.}~\bibnamefont {Kozio\l{}-Rachwa\l{}}}, \bibinfo {author}
  {\bibfnamefont {K.}~\bibnamefont {Matlak}}, \bibinfo {author} {\bibfnamefont
  {N.}~\bibnamefont {Spiridis}}, \bibinfo {author} {\bibfnamefont
  {D.}~\bibnamefont {Wilgocka-\ifmmode \acute{S}\else
  \'{S}\fi{}l\ifmmode~\mbox{\k{e}}\else \k{e}\fi{}zak}}, \bibinfo {author}
  {\bibfnamefont {E.}~\bibnamefont {Partyka-Jankowska}}, \bibinfo {author}
  {\bibfnamefont {M.}~\bibnamefont {Rennhofer}}, \bibinfo {author}
  {\bibfnamefont {A.~I.}\ \bibnamefont {Chumakov}}, \bibinfo {author}
  {\bibfnamefont {S.}~\bibnamefont {Stankov}}, \bibinfo {author} {\bibfnamefont
  {R.}~\bibnamefont {R\"uffer}},\ and\ \bibinfo {author} {\bibfnamefont
  {J.}~\bibnamefont {Korecki}},\ }\bibfield  {title} {\bibinfo {title}
  {Noncollinear magnetization structure at the thickness-driven
  spin-reorientation transition in epitaxial Fe films on W(110)},\ }\href
  {https://doi.org/10.1103/PhysRevLett.105.027206} {\bibfield  {journal}
  {\bibinfo  {journal} {Phys. Rev. Lett.}\ }\textbf {\bibinfo {volume} {105}},\
  \bibinfo {pages} {027206} (\bibinfo {year} {2010})}\BibitemShut {NoStop}%
\bibitem [{\citenamefont {Heide}\ \emph {et~al.}(2008)\citenamefont {Heide},
  \citenamefont {Bihlmayer},\ and\ \citenamefont {Bl\"ugel}}]{Heide2008}%
  \BibitemOpen
  \bibfield  {author} {\bibinfo {author} {\bibfnamefont {M.}~\bibnamefont
  {Heide}}, \bibinfo {author} {\bibfnamefont {G.}~\bibnamefont {Bihlmayer}},\
  and\ \bibinfo {author} {\bibfnamefont {S.}~\bibnamefont {Bl\"ugel}},\
  }\bibfield  {title} {\bibinfo {title} {Dzyaloshinskii-Moriya interaction
  accounting for the orientation of magnetic domains in ultrathin films:
  Fe/W(110)},\ }\href {https://doi.org/10.1103/PhysRevB.78.140403} {\bibfield
  {journal} {\bibinfo  {journal} {Phys. Rev. B}\ }\textbf {\bibinfo {volume}
  {78}},\ \bibinfo {pages} {140403(R)} (\bibinfo {year} {2008})}\BibitemShut
  {NoStop}%
\bibitem [{\citenamefont {Meckler}\ \emph {et~al.}(2009)\citenamefont
  {Meckler}, \citenamefont {Mikuszeit}, \citenamefont {Pre\ss{}ler},
  \citenamefont {Vedmedenko}, \citenamefont {Pietzsch},\ and\ \citenamefont
  {Wiesendanger}}]{Meckler2009}%
  \BibitemOpen
  \bibfield  {author} {\bibinfo {author} {\bibfnamefont {S.}~\bibnamefont
  {Meckler}}, \bibinfo {author} {\bibfnamefont {N.}~\bibnamefont {Mikuszeit}},
  \bibinfo {author} {\bibfnamefont {A.}~\bibnamefont {Pre\ss{}ler}}, \bibinfo
  {author} {\bibfnamefont {E.~Y.}\ \bibnamefont {Vedmedenko}}, \bibinfo
  {author} {\bibfnamefont {O.}~\bibnamefont {Pietzsch}},\ and\ \bibinfo
  {author} {\bibfnamefont {R.}~\bibnamefont {Wiesendanger}},\ }\bibfield
  {title} {\bibinfo {title} {Real-space observation of a right-rotating
  inhomogeneous cycloidal spin spiral by spin-polarized scanning tunneling
  microscopy in a triple axes vector magnet},\ }\href
  {https://doi.org/10.1103/PhysRevLett.103.157201} {\bibfield  {journal}
  {\bibinfo  {journal} {Phys. Rev. Lett.}\ }\textbf {\bibinfo {volume} {103}},\
  \bibinfo {pages} {157201} (\bibinfo {year} {2009})}\BibitemShut {NoStop}%
\bibitem [{\citenamefont {Udvardi}\ and\ \citenamefont
  {Szunyogh}(2009)}]{Udvardi}%
  \BibitemOpen
  \bibfield  {author} {\bibinfo {author} {\bibfnamefont {L.}~\bibnamefont
  {Udvardi}}\ and\ \bibinfo {author} {\bibfnamefont {L.}~\bibnamefont
  {Szunyogh}},\ }\bibfield  {title} {\bibinfo {title} {Chiral asymmetry of the
  spin-wave spectra in ultrathin magnetic films},\ }\href
  {https://doi.org/10.1103/PhysRevLett.102.207204} {\bibfield  {journal}
  {\bibinfo  {journal} {Phys. Rev. Lett.}\ }\textbf {\bibinfo {volume} {102}},\
  \bibinfo {pages} {207204} (\bibinfo {year} {2009})}\BibitemShut {NoStop}%
\bibitem [{\citenamefont {Fert}\ and\ \citenamefont {Levy}(1980)}]{Fert}%
  \BibitemOpen
  \bibfield  {author} {\bibinfo {author} {\bibfnamefont {A.}~\bibnamefont
  {Fert}}\ and\ \bibinfo {author} {\bibfnamefont {P.~M.}\ \bibnamefont
  {Levy}},\ }\bibfield  {title} {\bibinfo {title} {Role of anisotropic exchange
  interactions in determining the properties of spin-glasses},\ }\href
  {https://doi.org/10.1103/PhysRevLett.44.1538} {\bibfield  {journal} {\bibinfo
   {journal} {Phys. Rev. Lett.}\ }\textbf {\bibinfo {volume} {44}},\ \bibinfo
  {pages} {1538} (\bibinfo {year} {1980})}\BibitemShut {NoStop}%
\bibitem [{\citenamefont {Levy}\ and\ \citenamefont {Fert}(1981)}]{Levy}%
  \BibitemOpen
  \bibfield  {author} {\bibinfo {author} {\bibfnamefont {P.~M.}\ \bibnamefont
  {Levy}}\ and\ \bibinfo {author} {\bibfnamefont {A.}~\bibnamefont {Fert}},\
  }\bibfield  {title} {\bibinfo {title} {Anisotropy induced by nonmagnetic
  impurities in $\mathrm{Cu}$ Mn spin-glass alloys},\ }\href
  {https://doi.org/10.1103/PhysRevB.23.4667} {\bibfield  {journal} {\bibinfo
  {journal} {Phys. Rev. B}\ }\textbf {\bibinfo {volume} {23}},\ \bibinfo
  {pages} {4667} (\bibinfo {year} {1981})}\BibitemShut {NoStop}%
\bibitem [{\citenamefont {Cubukcu}\ \emph {et~al.}(2016)\citenamefont
  {Cubukcu}, \citenamefont {Sampaio}, \citenamefont {Bouzehouane},
  \citenamefont {Apalkov}, \citenamefont {Khvalkovskiy}, \citenamefont {Cros},\
  and\ \citenamefont {Reyren}}]{Cubukcu}%
  \BibitemOpen
  \bibfield  {author} {\bibinfo {author} {\bibfnamefont {M.}~\bibnamefont
  {Cubukcu}}, \bibinfo {author} {\bibfnamefont {J.}~\bibnamefont {Sampaio}},
  \bibinfo {author} {\bibfnamefont {K.}~\bibnamefont {Bouzehouane}}, \bibinfo
  {author} {\bibfnamefont {D.}~\bibnamefont {Apalkov}}, \bibinfo {author}
  {\bibfnamefont {A.~V.}\ \bibnamefont {Khvalkovskiy}}, \bibinfo {author}
  {\bibfnamefont {V.}~\bibnamefont {Cros}},\ and\ \bibinfo {author}
  {\bibfnamefont {N.}~\bibnamefont {Reyren}},\ }\bibfield  {title} {\bibinfo
  {title} {Dzyaloshinskii-Moriya anisotropy in nanomagnets with in-plane
  magnetization},\ }\href {https://doi.org/10.1103/PhysRevB.93.020401}
  {\bibfield  {journal} {\bibinfo  {journal} {Phys. Rev. B}\ }\textbf {\bibinfo
  {volume} {93}},\ \bibinfo {pages} {020401(R)} (\bibinfo {year}
  {2016})}\BibitemShut {NoStop}%
\bibitem [{\citenamefont {R\'ozsa}\ \emph {et~al.}(2017)\citenamefont
  {R\'ozsa}, \citenamefont {Atxitia},\ and\ \citenamefont {Nowak}}]{Rozsa}%
  \BibitemOpen
  \bibfield  {author} {\bibinfo {author} {\bibfnamefont {L.}~\bibnamefont
  {R\'ozsa}}, \bibinfo {author} {\bibfnamefont {U.}~\bibnamefont {Atxitia}},\
  and\ \bibinfo {author} {\bibfnamefont {U.}~\bibnamefont {Nowak}},\ }\bibfield
   {title} {\bibinfo {title} {Temperature scaling of the Dzyaloshinsky-Moriya
  interaction in the spin wave spectrum},\ }\href
  {https://doi.org/10.1103/PhysRevB.96.094436} {\bibfield  {journal} {\bibinfo
  {journal} {Phys. Rev. B}\ }\textbf {\bibinfo {volume} {96}},\ \bibinfo
  {pages} {094436} (\bibinfo {year} {2017})}\BibitemShut {NoStop}%
\bibitem [{\citenamefont {Yanes}\ \emph {et~al.}(2013)\citenamefont {Yanes},
  \citenamefont {Jackson}, \citenamefont {Udvardi}, \citenamefont {Szunyogh},\
  and\ \citenamefont {Nowak}}]{Yanes}%
  \BibitemOpen
  \bibfield  {author} {\bibinfo {author} {\bibfnamefont {R.}~\bibnamefont
  {Yanes}}, \bibinfo {author} {\bibfnamefont {J.}~\bibnamefont {Jackson}},
  \bibinfo {author} {\bibfnamefont {L.}~\bibnamefont {Udvardi}}, \bibinfo
  {author} {\bibfnamefont {L.}~\bibnamefont {Szunyogh}},\ and\ \bibinfo
  {author} {\bibfnamefont {U.}~\bibnamefont {Nowak}},\ }\bibfield  {title}
  {\bibinfo {title} {Exchange bias driven by Dzyaloshinskii-Moriya
  interactions},\ }\href {https://doi.org/10.1103/PhysRevLett.111.217202}
  {\bibfield  {journal} {\bibinfo  {journal} {Phys. Rev. Lett.}\ }\textbf
  {\bibinfo {volume} {111}},\ \bibinfo {pages} {217202} (\bibinfo {year}
  {2013})}\BibitemShut {NoStop}%
\bibitem [{\citenamefont {Fern\'{a}ndez-Pacheco}\ \emph
  {et~al.}(2019)\citenamefont {Fern\'{a}ndez-Pacheco}, \citenamefont
  {Vedmedenko}, \citenamefont {Ummelen}, \citenamefont {Mansell}, \citenamefont
  {Petit},\ and\ \citenamefont {Cowburn}}]{Fernandez-Pacheco2019}%
  \BibitemOpen
  \bibfield  {author} {\bibinfo {author} {\bibfnamefont {A.}~\bibnamefont
  {Fern\'{a}ndez-Pacheco}}, \bibinfo {author} {\bibfnamefont {E.}~\bibnamefont
  {Vedmedenko}}, \bibinfo {author} {\bibfnamefont {F.}~\bibnamefont {Ummelen}},
  \bibinfo {author} {\bibfnamefont {R.}~\bibnamefont {Mansell}}, \bibinfo
  {author} {\bibfnamefont {D.}~\bibnamefont {Petit}},\ and\ \bibinfo {author}
  {\bibfnamefont {R.~P.}\ \bibnamefont {Cowburn}},\ }\bibfield  {title}
  {\bibinfo {title} {Symmetry-breaking interlayer Dzyaloshinskii--Moriya
  interactions in synthetic antiferromagnets},\ }\href
  {https://doi.org/10.1038/s41563-019-0386-4} {\bibfield  {journal} {\bibinfo
  {journal} {Nat. Mater.}\ }\textbf {\bibinfo {volume} {18}},\ \bibinfo {pages}
  {679} (\bibinfo {year} {2019})}\BibitemShut {NoStop}%
\bibitem [{\citenamefont {Dashwood}\ \emph {et~al.}()\citenamefont {Dashwood},
  \citenamefont {Veiga}, \citenamefont {Faure}, \citenamefont {Vale},
  \citenamefont {Porter}, \citenamefont {Collins}, \citenamefont {Manuel},
  \citenamefont {Khalyavin}, \citenamefont {Orlandi}, \citenamefont {Perry},
  \citenamefont {Johnson},\ and\ \citenamefont {McMorrow}}]{Dashwood}%
  \BibitemOpen
  \bibfield  {author} {\bibinfo {author} {\bibfnamefont {C.~D.}\ \bibnamefont
  {Dashwood}}, \bibinfo {author} {\bibfnamefont {L.~S.~I.}\ \bibnamefont
  {Veiga}}, \bibinfo {author} {\bibfnamefont {Q.}~\bibnamefont {Faure}},
  \bibinfo {author} {\bibfnamefont {J.~G.}\ \bibnamefont {Vale}}, \bibinfo
  {author} {\bibfnamefont {D.~G.}\ \bibnamefont {Porter}}, \bibinfo {author}
  {\bibfnamefont {S.~P.}\ \bibnamefont {Collins}}, \bibinfo {author}
  {\bibfnamefont {P.}~\bibnamefont {Manuel}}, \bibinfo {author} {\bibfnamefont
  {D.~D.}\ \bibnamefont {Khalyavin}}, \bibinfo {author} {\bibfnamefont
  {F.}~\bibnamefont {Orlandi}}, \bibinfo {author} {\bibfnamefont {R.~S.}\
  \bibnamefont {Perry}}, \bibinfo {author} {\bibfnamefont {R.~D.}\ \bibnamefont
  {Johnson}},\ and\ \bibinfo {author} {\bibfnamefont {D.~F.}\ \bibnamefont
  {McMorrow}},\ }\href@noop {} {}\Eprint {https://arxiv.org/abs/2006.012882}
  {arXiv:2006.012882} \BibitemShut {NoStop}%
\bibitem [{\citenamefont {Udvardi}\ \emph {et~al.}(2003)\citenamefont
  {Udvardi}, \citenamefont {Szunyogh}, \citenamefont {Palot\'as},\ and\
  \citenamefont {Weinberger}}]{Udvardi2003}%
  \BibitemOpen
  \bibfield  {author} {\bibinfo {author} {\bibfnamefont {L.}~\bibnamefont
  {Udvardi}}, \bibinfo {author} {\bibfnamefont {L.}~\bibnamefont {Szunyogh}},
  \bibinfo {author} {\bibfnamefont {K.}~\bibnamefont {Palot\'as}},\ and\
  \bibinfo {author} {\bibfnamefont {P.}~\bibnamefont {Weinberger}},\ }\bibfield
   {title} {\bibinfo {title} {First-principles relativistic study of spin waves
  in thin magnetic films},\ }\href {https://doi.org/10.1103/PhysRevB.68.104436}
  {\bibfield  {journal} {\bibinfo  {journal} {Phys. Rev. B}\ }\textbf {\bibinfo
  {volume} {68}},\ \bibinfo {pages} {104436} (\bibinfo {year}
  {2003})}\BibitemShut {NoStop}%
\bibitem [{\citenamefont {Szunyogh}\ \emph {et~al.}(1995)\citenamefont
  {Szunyogh}, \citenamefont {\'Ujfalussy},\ and\ \citenamefont
  {Weinberger}}]{skkr}%
  \BibitemOpen
  \bibfield  {author} {\bibinfo {author} {\bibfnamefont {L.}~\bibnamefont
  {Szunyogh}}, \bibinfo {author} {\bibfnamefont {B.}~\bibnamefont
  {\'Ujfalussy}},\ and\ \bibinfo {author} {\bibfnamefont {P.}~\bibnamefont
  {Weinberger}},\ }\bibfield  {title} {\bibinfo {title} {Magnetic anisotropy of
  iron multilayers on Au(001): First-principles calculations in terms of the
  fully relativistic spin-polarized screened KKR method},\ }\href
  {https://doi.org/10.1103/PhysRevB.51.9552} {\bibfield  {journal} {\bibinfo
  {journal} {Phys. Rev. B}\ }\textbf {\bibinfo {volume} {51}},\ \bibinfo
  {pages} {9552} (\bibinfo {year} {1995})}\BibitemShut {NoStop}%
\bibitem [{\citenamefont {Albrecht}\ \emph {et~al.}(1991)\citenamefont
  {Albrecht}, \citenamefont {Gradmann}, \citenamefont {Reinert},\ and\
  \citenamefont {Fritsche}}]{Albrecht1991}%
  \BibitemOpen
  \bibfield  {author} {\bibinfo {author} {\bibfnamefont {M.}~\bibnamefont
  {Albrecht}}, \bibinfo {author} {\bibfnamefont {U.}~\bibnamefont {Gradmann}},
  \bibinfo {author} {\bibfnamefont {T.}~\bibnamefont {Reinert}},\ and\ \bibinfo
  {author} {\bibfnamefont {L.}~\bibnamefont {Fritsche}},\ }\bibfield  {title}
  {\bibinfo {title} {Scattering phases in low energy electron diffraction from
  spot profile analysis and from multiple scattering theory},\ }\href
  {https://doi.org/https://doi.org/10.1016/0038-1098(91)90399-G} {\bibfield
  {journal} {\bibinfo  {journal} {Sol. State Commun.}\ }\textbf {\bibinfo
  {volume} {78}},\ \bibinfo {pages} {671 } (\bibinfo {year}
  {1991})}\BibitemShut {NoStop}%
\bibitem [{\citenamefont {Tober}\ \emph {et~al.}(1997)\citenamefont {Tober},
  \citenamefont {Ynzunza}, \citenamefont {Palomares}, \citenamefont {Wang},
  \citenamefont {Hussain}, \citenamefont {Van~Hove},\ and\ \citenamefont
  {Fadley}}]{Tober1997}%
  \BibitemOpen
  \bibfield  {author} {\bibinfo {author} {\bibfnamefont {E.~D.}\ \bibnamefont
  {Tober}}, \bibinfo {author} {\bibfnamefont {R.~X.}\ \bibnamefont {Ynzunza}},
  \bibinfo {author} {\bibfnamefont {F.~J.}\ \bibnamefont {Palomares}}, \bibinfo
  {author} {\bibfnamefont {Z.}~\bibnamefont {Wang}}, \bibinfo {author}
  {\bibfnamefont {Z.}~\bibnamefont {Hussain}}, \bibinfo {author} {\bibfnamefont
  {M.~A.}\ \bibnamefont {Van~Hove}},\ and\ \bibinfo {author} {\bibfnamefont
  {C.~S.}\ \bibnamefont {Fadley}},\ }\bibfield  {title} {\bibinfo {title}
  {Interface structures of ordered Fe and Gd overlayers on W(110) from
  photoelectron diffraction},\ }\href
  {https://doi.org/10.1103/PhysRevLett.79.2085} {\bibfield  {journal} {\bibinfo
   {journal} {Phys. Rev. Lett.}\ }\textbf {\bibinfo {volume} {79}},\ \bibinfo
  {pages} {2085} (\bibinfo {year} {1997})}\BibitemShut {NoStop}%
\bibitem [{\citenamefont {Meyerheim}\ \emph {et~al.}(2001)\citenamefont
  {Meyerheim}, \citenamefont {Sander}, \citenamefont {Popescu}, \citenamefont
  {Kirschner}, \citenamefont {Steadman},\ and\ \citenamefont
  {Ferrer}}]{Meyerheim2001}%
  \BibitemOpen
  \bibfield  {author} {\bibinfo {author} {\bibfnamefont {H.~L.}\ \bibnamefont
  {Meyerheim}}, \bibinfo {author} {\bibfnamefont {D.}~\bibnamefont {Sander}},
  \bibinfo {author} {\bibfnamefont {R.}~\bibnamefont {Popescu}}, \bibinfo
  {author} {\bibfnamefont {J.}~\bibnamefont {Kirschner}}, \bibinfo {author}
  {\bibfnamefont {P.}~\bibnamefont {Steadman}},\ and\ \bibinfo {author}
  {\bibfnamefont {S.}~\bibnamefont {Ferrer}},\ }\bibfield  {title} {\bibinfo
  {title} {Surface structure and stress in Fe monolayers on W(110)},\ }\href
  {https://doi.org/10.1103/PhysRevB.64.045414} {\bibfield  {journal} {\bibinfo
  {journal} {Phys. Rev. B}\ }\textbf {\bibinfo {volume} {64}},\ \bibinfo
  {pages} {045414} (\bibinfo {year} {2001})}\BibitemShut {NoStop}%
\bibitem [{\citenamefont {Qian}\ and\ \citenamefont
  {H\"ubner}(1999)}]{Qian1999}%
  \BibitemOpen
  \bibfield  {author} {\bibinfo {author} {\bibfnamefont {X.}~\bibnamefont
  {Qian}}\ and\ \bibinfo {author} {\bibfnamefont {W.}~\bibnamefont
  {H\"ubner}},\ }\bibfield  {title} {\bibinfo {title} {First-principles
  calculation of structural and magnetic properties for Fe monolayers and
  bilayers on W(110)},\ }\href {https://doi.org/10.1103/PhysRevB.60.16192}
  {\bibfield  {journal} {\bibinfo  {journal} {Phys. Rev. B}\ }\textbf {\bibinfo
  {volume} {60}},\ \bibinfo {pages} {16192} (\bibinfo {year}
  {1999})}\BibitemShut {NoStop}%
\bibitem [{\citenamefont {Santos}\ \emph {et~al.}(2016)\citenamefont {Santos},
  \citenamefont {Rybicki}, \citenamefont {Zasada}, \citenamefont {Starodub},
  \citenamefont {McCarty}, \citenamefont {Cerda}, \citenamefont {Puerta},\ and\
  \citenamefont {de~la Figuera}}]{Santos2016}%
  \BibitemOpen
  \bibfield  {author} {\bibinfo {author} {\bibfnamefont {B.}~\bibnamefont
  {Santos}}, \bibinfo {author} {\bibfnamefont {M.}~\bibnamefont {Rybicki}},
  \bibinfo {author} {\bibfnamefont {I.}~\bibnamefont {Zasada}}, \bibinfo
  {author} {\bibfnamefont {E.}~\bibnamefont {Starodub}}, \bibinfo {author}
  {\bibfnamefont {K.~F.}\ \bibnamefont {McCarty}}, \bibinfo {author}
  {\bibfnamefont {J.~I.}\ \bibnamefont {Cerda}}, \bibinfo {author}
  {\bibfnamefont {J.~M.}\ \bibnamefont {Puerta}},\ and\ \bibinfo {author}
  {\bibfnamefont {J.}~\bibnamefont {de~la Figuera}},\ }\bibfield  {title}
  {\bibinfo {title} {Structure and stability of ultrathin Fe films on W(110)},\
  }\href {https://doi.org/10.1103/PhysRevB.93.195423} {\bibfield  {journal}
  {\bibinfo  {journal} {Phys. Rev. B}\ }\textbf {\bibinfo {volume} {93}},\
  \bibinfo {pages} {195423} (\bibinfo {year} {2016})}\BibitemShut {NoStop}%
\bibitem [{\citenamefont {Heide}(2006)}]{Heide2006}%
  \BibitemOpen
  \bibfield  {author} {\bibinfo {author} {\bibfnamefont {M.}~\bibnamefont
  {Heide}},\ }\emph {\bibinfo {title} {Magnetic domain walls in ultrathin
  films: Contribution of the Dzyaloshinsky-Moriya interaction}},\ \href@noop {}
  {Ph.D. thesis},\ \bibinfo  {school} {Rheinisch-Westfälische Technische
  Hochschule Aachen} (\bibinfo {year} {2006})\BibitemShut {NoStop}%
\bibitem [{sup()}]{supp}%
  \BibitemOpen
  \href@noop {} {}\bibinfo {note} {See Supplemental Material at [URL] for the tensorial interaction parameters in the Fe DL on
  W(110).}\BibitemShut {Stop}%
\bibitem [{\citenamefont {{Laio}}\ and\ \citenamefont
  {{Parrinello}}(2002)}]{Laio2002}%
  \BibitemOpen
  \bibfield  {author} {\bibinfo {author} {\bibfnamefont {A.}~\bibnamefont
  {{Laio}}}\ and\ \bibinfo {author} {\bibfnamefont {M.}~\bibnamefont
  {{Parrinello}}},\ }\bibfield  {title} {\bibinfo {title} {{Escaping
  free-energy minima}},\ }\href {https://doi.org/10.1073/pnas.202427399}
  {\bibfield  {journal} {\bibinfo  {journal} {Proc. Nat. Acad. Sci.}\ }\textbf
  {\bibinfo {volume} {99}},\ \bibinfo {pages} {12562} (\bibinfo {year}
  {2002})}\BibitemShut {NoStop}%
\bibitem [{\citenamefont {{Barducci}}\ \emph {et~al.}(2008)\citenamefont
  {{Barducci}}, \citenamefont {{Bussi}},\ and\ \citenamefont
  {{Parrinello}}}]{Barducci2008}%
  \BibitemOpen
  \bibfield  {author} {\bibinfo {author} {\bibfnamefont {A.}~\bibnamefont
  {{Barducci}}}, \bibinfo {author} {\bibfnamefont {G.}~\bibnamefont
  {{Bussi}}},\ and\ \bibinfo {author} {\bibfnamefont {M.}~\bibnamefont
  {{Parrinello}}},\ }\bibfield  {title} {\bibinfo {title} {{Well-Tempered
  Metadynamics: A Smoothly Converging and Tunable Free-Energy Method}},\ }\href
  {https://doi.org/10.1103/PhysRevLett.100.020603} {\bibfield  {journal}
  {\bibinfo  {journal} {Phys. Rev. Lett.}\ }\textbf {\bibinfo {volume} {100}},\
  \bibinfo {eid} {020603} (\bibinfo {year} {2008})}\BibitemShut {NoStop}%
\bibitem [{\citenamefont {Nagyfalusi}\ \emph {et~al.}(2019)\citenamefont
  {Nagyfalusi}, \citenamefont {Udvardi},\ and\ \citenamefont
  {Szunyogh}}]{Nagyfalusi2019}%
  \BibitemOpen
  \bibfield  {author} {\bibinfo {author} {\bibfnamefont {B.}~\bibnamefont
  {Nagyfalusi}}, \bibinfo {author} {\bibfnamefont {L.}~\bibnamefont
  {Udvardi}},\ and\ \bibinfo {author} {\bibfnamefont {L.}~\bibnamefont
  {Szunyogh}},\ }\bibfield  {title} {\bibinfo {title} {Metadynamics study of
  the temperature dependence of magnetic anisotropy and spin-reorientation
  transitions in ultrathin films},\ }\href
  {https://doi.org/10.1103/PhysRevB.100.174429} {\bibfield  {journal} {\bibinfo
   {journal} {Phys. Rev. B}\ }\textbf {\bibinfo {volume} {100}},\ \bibinfo
  {pages} {174429} (\bibinfo {year} {2019})}\BibitemShut {NoStop}%
\bibitem [{\citenamefont {\ifmmode \acute{S}\else
  \'{S}\fi{}l\ifmmode~\mbox{\k{e}}\else \k{e}\fi{}zak}\ \emph
  {et~al.}(2013)\citenamefont {\ifmmode \acute{S}\else
  \'{S}\fi{}l\ifmmode~\mbox{\k{e}}\else \k{e}\fi{}zak}, \citenamefont {\ifmmode
  \acute{S}\else \'{S}\fi{}l\ifmmode~\mbox{\k{e}}\else \k{e}\fi{}zak},
  \citenamefont {Freindl}, \citenamefont {Kara\ifmmode~\acute{s}\else
  \'{s}\fi{}}, \citenamefont {Spiridis}, \citenamefont {Zaj\k{a}c},
  \citenamefont {Chumakov}, \citenamefont {Stankov}, \citenamefont {R\"uffer},\
  and\ \citenamefont {Korecki}}]{Slezak2013}%
  \BibitemOpen
  \bibfield  {author} {\bibinfo {author} {\bibfnamefont {M.}~\bibnamefont
  {\ifmmode \acute{S}\else \'{S}\fi{}l\ifmmode~\mbox{\k{e}}\else
  \k{e}\fi{}zak}}, \bibinfo {author} {\bibfnamefont {T.}~\bibnamefont {\ifmmode
  \acute{S}\else \'{S}\fi{}l\ifmmode~\mbox{\k{e}}\else \k{e}\fi{}zak}},
  \bibinfo {author} {\bibfnamefont {K.}~\bibnamefont {Freindl}}, \bibinfo
  {author} {\bibfnamefont {W.}~\bibnamefont {Kara\ifmmode~\acute{s}\else
  \'{s}\fi{}}}, \bibinfo {author} {\bibfnamefont {N.}~\bibnamefont {Spiridis}},
  \bibinfo {author} {\bibfnamefont {M.}~\bibnamefont {Zaj\k{a}c}}, \bibinfo
  {author} {\bibfnamefont {A.~I.}\ \bibnamefont {Chumakov}}, \bibinfo {author}
  {\bibfnamefont {S.}~\bibnamefont {Stankov}}, \bibinfo {author} {\bibfnamefont
  {R.}~\bibnamefont {R\"uffer}},\ and\ \bibinfo {author} {\bibfnamefont
  {J.}~\bibnamefont {Korecki}},\ }\bibfield  {title} {\bibinfo {title}
  {Perpendicular magnetic anisotropy and noncollinear magnetic structure in
  ultrathin Fe films on W(110)},\ }\href
  {https://doi.org/10.1103/PhysRevB.87.134411} {\bibfield  {journal} {\bibinfo
  {journal} {Phys. Rev. B}\ }\textbf {\bibinfo {volume} {87}},\ \bibinfo
  {pages} {134411} (\bibinfo {year} {2013})}\BibitemShut {NoStop}%
\bibitem [{\citenamefont {Evans}\ \emph {et~al.}()\citenamefont {Evans},
  \citenamefont {R\'{o}zsa}, \citenamefont {Jenkins},\ and\ \citenamefont
  {Atxitia}}]{Evans2020}%
  \BibitemOpen
  \bibfield  {author} {\bibinfo {author} {\bibfnamefont {R.~F.~L.}\
  \bibnamefont {Evans}}, \bibinfo {author} {\bibfnamefont {L.}~\bibnamefont
  {R\'{o}zsa}}, \bibinfo {author} {\bibfnamefont {S.}~\bibnamefont {Jenkins}},\
  and\ \bibinfo {author} {\bibfnamefont {U.}~\bibnamefont {Atxitia}},\
  }\href@noop {} {}\bibfield
  {title} {\bibinfo {title} {Temperature scaling of two-ion anisotropy in pure and mixed anisotropy systems},\
  }\href {https://doi.org/10.1103/PhysRevB.102.020412}{\bibfield  {journal} {\bibinfo  {journal} {Phys. Rev. B}\ }\textbf {\bibinfo
  {volume} {102}},\ \bibinfo {pages} {020412(R)} (\bibinfo {year}
  {2020})} \BibitemShut {NoStop}%
\bibitem [{\citenamefont {Han}\ \emph {et~al.}(2019)\citenamefont {Han},
  \citenamefont {Lee}, \citenamefont {Hanke}, \citenamefont {Mokrousov},
  \citenamefont {Kim}, \citenamefont {Yoo}, \citenamefont {van Hees},
  \citenamefont {Kim}, \citenamefont {Lavrijsen}, \citenamefont {You},
  \citenamefont {Swagten}, \citenamefont {Jung},\ and\ \citenamefont
  {Kl\"{a}ui}}]{Han2019}%
  \BibitemOpen
  \bibfield  {author} {\bibinfo {author} {\bibfnamefont {D.-S.}\ \bibnamefont
  {Han}}, \bibinfo {author} {\bibfnamefont {K.}~\bibnamefont {Lee}}, \bibinfo
  {author} {\bibfnamefont {J.-P.}\ \bibnamefont {Hanke}}, \bibinfo {author}
  {\bibfnamefont {Y.}~\bibnamefont {Mokrousov}}, \bibinfo {author}
  {\bibfnamefont {K.-W.}\ \bibnamefont {Kim}}, \bibinfo {author} {\bibfnamefont
  {W.}~\bibnamefont {Yoo}}, \bibinfo {author} {\bibfnamefont {Y.~L.~W.}\
  \bibnamefont {van Hees}}, \bibinfo {author} {\bibfnamefont {T.-W.}\
  \bibnamefont {Kim}}, \bibinfo {author} {\bibfnamefont {R.}~\bibnamefont
  {Lavrijsen}}, \bibinfo {author} {\bibfnamefont {C.-Y.}\ \bibnamefont {You}},
  \bibinfo {author} {\bibfnamefont {H.~J.~M.}\ \bibnamefont {Swagten}},
  \bibinfo {author} {\bibfnamefont {M.-H.}\ \bibnamefont {Jung}},\ and\
  \bibinfo {author} {\bibfnamefont {M.}~\bibnamefont {Kl\"{a}ui}},\ }\bibfield
  {title} {\bibinfo {title} {Long-range chiral exchange interaction in
  synthetic antiferromagnets},\ }\href
  {https://doi.org/10.1038/s41563-019-0370-z} {\bibfield  {journal} {\bibinfo
  {journal} {Nat. Mater.}\ }\textbf {\bibinfo {volume} {18}},\ \bibinfo {pages}
  {703} (\bibinfo {year} {2019})}\BibitemShut {NoStop}%
\bibitem [{\citenamefont {Vedmedenko}\ \emph {et~al.}(2019)\citenamefont
  {Vedmedenko}, \citenamefont {Riego}, \citenamefont {Arregi},\ and\
  \citenamefont {Berger}}]{Vedmedenko2019}%
  \BibitemOpen
  \bibfield  {author} {\bibinfo {author} {\bibfnamefont {E.~Y.}\ \bibnamefont
  {Vedmedenko}}, \bibinfo {author} {\bibfnamefont {P.}~\bibnamefont {Riego}},
  \bibinfo {author} {\bibfnamefont {J.~A.}\ \bibnamefont {Arregi}},\ and\
  \bibinfo {author} {\bibfnamefont {A.}~\bibnamefont {Berger}},\ }\bibfield
  {title} {\bibinfo {title} {Interlayer Dzyaloshinskii-Moriya interactions},\
  }\href {https://doi.org/10.1103/PhysRevLett.122.257202} {\bibfield  {journal}
  {\bibinfo  {journal} {Phys. Rev. Lett.}\ }\textbf {\bibinfo {volume} {122}},\
  \bibinfo {pages} {257202} (\bibinfo {year} {2019})}\BibitemShut {NoStop}%
\end{thebibliography}

%

\end{document}